\documentclass[10pt]{article}
\usepackage{graphicx,floatflt,amssymb,epsfig,epsf} 
\def\be {\begin{equation}}
\def\ee {\end{equation}}
\def\bea {\begin{eqnarray}}
\def\eea {\end{eqnarray}}

\def\bra {\langle}
\def\ket {\rangle}
                                                                                
\def\g {\gamma}

\def\l {\lambda}

\def\bbar {\overline{B}{}^0}

\def\bbbar {B^0-\bbar}

\def\bsss {b\to s\bar{s} s}
\def\bccs {b\to c\bar{c} s}
\def\bsbsbar {B_s-\overline{B_s}}

\def\opcit(#1){ {\em op. cit.}, #1}
\def\etal {\em et al.}

\def\issue(#1,#2,#3){#1 (#3) #2} 

\def\APP(#1,#2,#3){Acta Phys.\ Polon.\ \issue(#1,#2,#3)}
\def\ARNPS(#1,#2,#3){Ann.\ Rev.\ Nucl.\ Part.\ Sci.\ \issue(#1,#2,#3)}
\def\CPC(#1,#2,#3){Comp.\ Phys.\ Comm.\ \issue(#1,#2,#3)}
\def\CIP(#1,#2,#3){Comput.\ Phys.\ \issue(#1,#2,#3)}
\def\EPJC(#1,#2,#3){Eur.\ Phys.\ J.\ C\ \issue(#1,#2,#3)}
\def\EPJD(#1,#2,#3){Eur.\ Phys.\ J. Direct\ C\ \issue(#1,#2,#3)}
\def\IEEETNS(#1,#2,#3){IEEE Trans.\ Nucl.\ Sci.\ \issue(#1,#2,#3)}
\def\IJMP(#1,#2,#3){Int.\ J.\ Mod.\ Phys. \issue(#1,#2,#3)}
\def\JHEP(#1,#2,#3){J.\ High Energy Physics \issue(#1,#2,#3)}
\def\JPG(#1,#2,#3){J.\ Phys.\ G \issue(#1,#2,#3)}
\def\MPL(#1,#2,#3){Mod.\ Phys.\ Lett.\ \issue(#1,#2,#3)}
\def\NP(#1,#2,#3){Nucl.\ Phys.\ \issue(#1,#2,#3)}
\def\NIM(#1,#2,#3){Nucl.\ Instrum.\ Meth.\ \issue(#1,#2,#3)}
\def\PL(#1,#2,#3){Phys.\ Lett.\ \issue(#1,#2,#3)}
\def\PRD(#1,#2,#3){Phys.\ Rev.\ D \issue(#1,#2,#3)}
\def\PRL(#1,#2,#3){Phys.\ Rev.\ Lett.\ \issue(#1,#2,#3)}
\def\SJNP(#1,#2,#3){Sov.\ J. Nucl.\ Phys.\ \issue(#1,#2,#3)}
\def\ZPC(#1,#2,#3){Zeit.\ Phys.\ C \issue(#1,#2,#3)}

\begin{document} 
\begin{flushright} 
CU-PHYSICS/08-2005\\
\end{flushright} 
\vskip 30pt 
 
\begin{center} 
{\Large \bf New Physics in \boldmath $b \to s \bar{s} s$\unboldmath~ Decay}\\
\vspace*{1cm} 
\renewcommand{\thefootnote}{\fnsymbol{footnote}} 
{\large {\sf Anirban Kundu ${}^1$}, {\sf Soumitra Nandi ${}^1$}, 
and {\sf Jyoti Prasad Saha ${}^2$} 
} \\ 
\vspace{10pt} 
{\small 
   ${}^{1)}$ {\em Department of Physics, University of Calcutta, 92 A.P.C. 
        Road, Kolkata 700009, India}\\  
   ${}^{2)}$ {\em Department of Physics, Jadavpur University, 
        Kolkata - 700032, India}} 
 
\normalsize 
\end{center} 
 
\begin{abstract} 
We perform a model-independent analysis of the data on branching ratios and
CP asymmetries of $B\to\phi K$ and $B\to\eta^{(')} K^{(*)}$ modes.
The present data is encouraging to look for indirect evidences of physics
beyond the Standard Model. We investigate
the parameter spaces for different possible Lorentz structures
of the new physics four-Fermi interaction. It is shown that if one takes
the data at $1\sigma$ confidence level, only one particular Lorentz 
structure is allowed. Possible consequences for the $B_s$ system are 
also discussed. 
 
\vskip 5pt \noindent 
\texttt{PACS numbers:~ 13.25.Hw, 14.40.Nd, 12.15.Hh} \\ 
\texttt{Keywords:~~B Decays, B$_{\tt s}$ Meson, Physics Beyond Standard Model}
\end{abstract}

\renewcommand{\thesection}{\Roman{section}} 
\setcounter{footnote}{0} 
\renewcommand{\thefootnote}{\arabic{footnote}} 

\section{Introduction}

The data from B factories has reached a state where one can talk about
looking for indirect signals of physics beyond the Standard Model (BSM)
\cite{hfag,utfit}. 
Before the Large Hadron Collider (LHC) comes up, it seems we need a fare
share of luck to directly produce BSM particles, be it of supersymmetry,
technicolour, compactified extra dimensions, models with extra chiral/vector 
fermions, gauge bosons or scalars, or of some other extensions of the Standard
Model (SM). On the other hand, radiative corrections induced by such new
particles may affect low-energy observables. 

The effects of specific models of BSM on low-energy data, particularly data
from B factories, have been extensively discussed in the literature.
Indeed, there is some justification for such an enthusiasm. It seems that the
data is not exactly in harmony with the SM expectations. There are two
caveats: the error bars are still large enough, and the low-energy dynamics
is not very well known. The second point can be tackled by looking at the
mixing-induced CP asymmetry data, where hadronic uncertainties mostly
cancel out. Unfortunately, only for a limited number of channels such 
asymmetries have been measured. It is hoped that in near future we may
have a situation where the error in the data will be dominated by theoretical
uncertainties. 
We can only hope that the experimental numbers that we now have 
will stand the test of time and proceed accordingly.

If one looks carefully, the anomalous results seem to be concentrated in the
$\bsss$ decay sector. One can say that this is apparently so because we
do not really understand the penguin dynamics, but the radiative decay
$b\to s\gamma$ is well explained within the SM.
This has prompted a number of analyses of the nonleptonic data, particularly
those of $B\to\phi K$, in the
context of specific new physics models, like different versions of
supersymmetry \cite{bphik:susy}, and non-supersymmetric models like extra
$Z'$, more Higgs doublets, extra fermions, etc. \cite{bphik:nonsusy}.
A good guide about the anomaly is the Heavy Flavor Averaging Group (HFAG)
website \cite{hfag} (also the UTfit collaboration \cite{utfit} performed
some model-independent new physics analyses). 
The value of $\sin(2\beta)$, as extracted from all
charmonium modes, is $0.725\pm 0.037$ (this is consistent with the value
obtained from CKM fits without superimposing the direct CP asymmetry
measurement data), while it drops down to $0.43\pm 0.07$
for $\bsss$ modes. Both sets are internally consistent as far as exclusive
modes are concerned. However, there is a second important factor: the branching
ratios (BR) for $B\to \eta' K$ and $B\to \eta K^\ast$ channels are
abnormally large, at least if we do not take into account the {\em a
posteriori} explanations of a large charm content of $\eta$ or $\eta'$ or
the significant gluonic contributions.
Some of the papers in \cite{bphik:susy} also deal with this BR enhancement
problem. 
The data is shown in table 1.

Still, the results are not inconsistent with the SM. If we look at
table 1, it is seen that $\sin(2\beta)$ extracted from $B\to\phi K_S$ is
consistent with that extracted from the charmonium modes at less than $2\sigma$.
The BR data is never a clear indicator; {\em e.g.}, in
the factorisation model \cite{ali} the $B\to\phi K$ BR is
quite unstable with the variation of $N_c$, the effective number of colours,
and one can always take help from SM dynamics beyond naive valence quark
model to explain the BRs of 
$\eta(')$ modes. Only the value of $\sin(2\beta)$ from
$B\to\eta' K$ is away from the charmonium value by almost $3\sigma$. It is
imperative to have more precise values of the CP asymmetries shown in table
1, and also accurate CP asymmetries from channels like $B\to 3K_S$
or $B\to K^+K^-K_S$ \cite{babar-3k}.

Let us assume that the data, as it is, will stand the test of time. Even then
this does not mean that only the $\bsss$ channel is affected by BSM physics.
Indeed, it will be very difficult to find a model which affects only one
single channel. In particular, if at least two of the final state quarks
are left-handed, then the SU(2) conjugate channel, $\bccs$, will be there.
The deciding factor is the SM contribution: for the former, it is a 
penguin contribution, whereas for the latter, it is a tree-level one, so
the BSM amplitude has less chance to compete and show up in $\bccs$ decays. 

In this letter we perform a model-independent study of all $\bsss$ channels
that have shown signs of anomaly. There are a number of works in the literature
\cite{bphik:modind,superb} 
that stress the need for such a model-independent analysis.
We investigate what type of Lorentz structure
of the four-Fermi interaction coming from the BSM physics can simultaneously
solve all the anomalies. For simplicity, we do not include tensor currents,
but all other bilinear covariants are taken into account. This serves as
a testing ground for BSM models as far as their ability in explaining the
anomalous B decay results are concerned. We will also assume that the 
new physics, whatever it is, does not contribute significantly to $\bbbar$
mixing; in particular, to the CP asymmetry prediction from the $\bbbar$
box diagram ({\em e.g.}, minimal supersymmetry with perfect alignment, {\em
i.e.}, all flavour-changing $\delta$ parameters set to zero). 

We will show that it is impossible to find a simultaneous solution for
all BRs and CP asymmetries with a hierarchical structure of new physics
({\em i.e.}, only one Lorentz structure is numerically significant). 
The reason has been explained in section 3.
We find that it is necessary to
have some dynamics within SM but beyond the naive valence quark model
(NVQM) to explain the BRs of the $\eta(')$ modes. Even with such an
assumption, we have a fairly constrained parameter space, different for
structurally different four-Fermi interactions, which can be further tested
in hadronic machines (now that BTeV has been cancelled, this essentially
means LHC-B). 

Theoretically, a new Lorentz structure as discussed above should contribute
to $\bsbsbar$ mixing. This can potentially change the weak phase 
coming from the $\bsbsbar$ box from the SM prediction of almost
zero. This will directly affect the measured CP asymmeries in $B_s\to J/\psi
\phi$, $B_s\to D_s^+ D_s^-$, $B_s\to \eta^{(')}\eta^{(')}$ channels. 
Unfortunately, the allowed parameter space for new physics can contribute
only marginally in $\bsbsbar$ mixing; the SM contribution, whose lower limit
only exists, is too large. Thus, the effects can only be studied (hopefully)
in a super-B factory, an $e^+e^-$ collider sitting on the $\Upsilon(5S)$ 
resonance.

The first two decay channels, {\em i.e.}, $B_s\to J/\psi \phi$ and
$B_s\to D_s^+ D_s^-$ will be further affected from BSM physics in decay
if an SU(2) partner $\bccs$ channel is present. However, we will show that
this effect can be neglected. On the other hand, the channels
$B_s\to \phi\phi$ or $B_s\to \eta^{(')}\eta^{(')}$, controlled
by $\bsss$ transition, should show modification from the SM prediction on two
grounds: BSM physics in mixing as well as in decay. These predictions will
be quantitatively different for different Lorentz structures.

\section{Data, Theory, and New Physics}

The relevant data, after Moriond 2005, is taken from the
HFAG website \cite{hfag} (updated April 2005) and is shown in table 1.
\begin{table}[htbp]
\begin{center}
\begin{tabular}{|l|c|c|c|}
\hline
Final state & BR $\times 10^6$  &  $\sin(2\beta)$  &  $A_{CP}^{dir}$\\
\hline
$\phi K_S$        &  $8.3 ^{+1.2}_{-1.0 }$  &  $0.34\pm 0.20$ &$-0.04 \pm 0.17$
        \\
$\eta K^{0\ast}$  &  $18.7 \pm 1.7$         & & $0.01\pm 0.08$ \\
$\eta' K_S$       &  $68.6 \pm 4.2$  &  $0.43\pm 0.11$ &$-0.04 \pm 0.08$
        \\
$\eta' K^{0\ast}$ &  $< 7.6 $               & & \\
$\eta K_S$        &  $< 2.0 $         & & \\
$\phi K^+$        &  $9.03^{+0.65}_{-0.63}$ & & $-0.037 \pm 0.050$ \\
$\eta K^{+\ast}$  &  $24.3^{+3.0}_{-2.9}$   & & $-0.03 ^{+0.10}_{-0.11}$ \\
$\eta' K^+$       &  $70.8 \pm 3.4 $  & & $-0.027\pm 0.025$ \\
$\eta' K^{+\ast}$ &  $< 14$                 & & \\
$\eta K^+$        &  $2.6\pm 0.5$           & & $0.25\pm 0.14$  \\
\hline
\end{tabular}
\caption{Experimental data on $\bsss$ modes. For our convention of
$A_{CP}$, see text. The third column shows the value of $\sin(2\beta)$ as
extracted from mixing-induced CP asymmetry of the corresponding modes.
The error bars are at $1\sigma$ confidence limit (CL); the upper limits are
at 90\% CL.} 
\end{center}
\end{table}

The theoretical numbers, particularly those for the BRs,
have some inherent uncertainties that stem from low-energy QCD. We use the
numbers from the conventional factorisation model with $N_c=3$ \cite{ali,bsw}.
However, one may question the validity of such a simplified approach. Other
models like QCD factorisation \cite{beneke} or perturbative QCD \cite{sanda,
keum} differ in their predictions of the BRs for these modes. The reason
for this difference is mainly twofold: first, the form factors for the
factorizable diagrams are different, and second, the nonfactorizable
topologies (like annihilation or emission) are given unequal importance
in these models. A study shows that the predictions for the amplitudes
vary at most by 20\% ({\em i.e.}, more than 40\% variation in the BRs)
for the $N_c$-stable modes like $\eta(')K$, while $\phi K$ modes are
rather unstable and the amplitude uncertainty can go up to 40\%.
Since there is no concensus about which model one should use, we take
the conventional factorisation numbers with an error margin of 20\% in
the amplitude level (and 40\% for $\phi K$ modes). This is just to
take into account the variation of predictions for different models
and should not be confused with the variation of a model parameter,
say, $N_c$.  
The charm contribution in $\eta$ or $\eta'$ is neglected to start with,
since that is an {\em a posteriori} approach to fix the BRs with the experiment.
However, we will soon see that one must entertain either the possibility of
a significant charm content in these mesons or a large gluonic contribution,
{\em i.e.}, something beyond the NVQM, to satisfy the data even in the 
presence of new physics.

The CP asymmetries, particularly the mixing-induced ones that measure
$\sin(2\beta)$, are more or less free from theoretical uncertainties. 
We assume that charm and up penguins are negligible compared to the
top penguin, so that in the SM, $B\to\phi K_S$ is essentially a one-amplitude
process. (Again, one can assume some phenomenological values for these
penguins which reproduce the CP asymmetry data, thereby alleviating the
need for any new physics.) The CP asymmetries are defined as
\be
A_{CP}=\frac{\Gamma(B^+/B^0 \to f)-\Gamma(B^-/\bbar \to \bar{f})}
{\Gamma(B^+/B^0 \to f)+\Gamma(B^-/\bbar \to \bar{f})},
\ee
which differs in sign from the HFAG convention. More precisely, defining
\be
\l=e^{-i\phi_M}{\bra f|\bar{B_d}\ket\over \bra f|{B_d}\ket},
\ee
where $\phi_M$ is the mixing phase ($\phi_M=2\beta$ for $\bbbar$ mixing
in the SM), we have
\be
a^d_{CP}={1-|\l|^2\over 1+|\l|^2}; \ \
a^m_{CP}={2Im\l\over 1+|\l|^2},
\ee
and the conventional $S$ and $C (A)$ parameters are given by
\be
S=-a^m_{CP}, \ \ C=-A=a^d_{CP}.
\ee

For numerical analysis, we follow the values given in \cite{ali}; 
{\em e.g.}, the decay constants (in GeV) are
\be
f_\pi = 0.133,\ \ f_K=0.158,\ \ f_\phi = 0.233,
\ee
and we use $f_1=1.10f_\pi$, $f_8=1.34 f_\pi$ with the one-angle mixing scheme
(the two-angle one \cite{escribano} makes little difference)
\bea
f^u_\eta=\frac{f_8 \cos\theta}{\sqrt{6}}-\frac{f_1 \sin\theta}{\sqrt{3}},\ &{}&
f^s_\eta=-2\frac{f_8 \cos\theta}{\sqrt{6}}-\frac{f_1 \sin\theta}{\sqrt{3}},
\nonumber\\
f^u_{\eta'}=\frac{f_8 \sin\theta}{\sqrt{6}}+\frac{f_1 \cos\theta}{\sqrt{3}},
\ &{}&
f^s_{\eta'}=-2\frac{f_8 \sin\theta}{\sqrt{6}}+\frac{f_1 \cos\theta}{\sqrt{3}}.
\eea
Here $\theta=-22^\circ$, and
$if^u_\eta p^\mu_\eta=\bra \eta|\bar{u}\g^\mu (1-\g_5) u|0\ket$, and similarly
for other decay constants. 
The numerical values are
$f^u_\eta=0.099$, $f^s_\eta=-0.103$, $f^u_{\eta'}=0.051$, $f^s_{\eta'}=0.133$.
Note that $f^s_\eta$ and $f^s_{\eta'}$ are both large and of the same order
of magnitude; this will be relevant for our future analysis.

The expressions for the SM amplitudes can be found in \cite{ali}. The
regularisation scale is taken to be $\mu=m_b/2\approx 2.5$ GeV 
and the quark masses according to \cite{ali} ($m_s=0.5$ GeV, $m_u=m_d=
0.2$ GeV). We
use the Wilson coefficients evaluated for $N_c=3$ at next-to-leading order
(NLO). Considering the fact
that we have considered some inherent uncertainty in the amplitudes, it
is expected that the results will be more or less
stable with the variation of $N_c$. 
This is indeed found to be the case, but for more comments the reader is
referred to the next section.

We discuss two different types of effective four-Fermi interactions
coming from new physics:
\bea
{\rm 1. ~Scalar:} \ \ &{}&
{\cal L}_{new} = h_s e^{i\phi_s}
\left( \bar{s}_{\alpha} (c_1+c_2\g_5) s_\alpha\right)
\left( \bar{s}_{\beta} (c_3+c_4\g_5) b_\beta\right),\nonumber\\
{\rm 2.~ Vector:} \ \ &{}&
{\cal L}_{new} = h_v e^{i\phi_v}
\left( \bar{s}_{\alpha} \g^\mu(c_1+c_2\g_5) s_\alpha\right)
\left( \bar{s}_{\beta} \g_\mu (c_3+c_4\g_5) b_\beta\right).
\eea
Here $\alpha$ and $\beta$ are colour indices. The couplings $h_{s,v}$
are effective couplings, of dimension $[M]^{-2}$, that one obtains by
integrating out the new physics fields. They are assumed to be real and
positive and the weak phase information is dumped in the quantities
$\phi_{s,v}$, which can vary in the range 0-$2\pi$. 
Note that they are effective couplings at the {\em weak} scale,
which one may obtain by incorporating all RG effects to the high-scale
values of them. 
The couplings $c_1$-$c_4$ can take any values between
$-1$ and 1; to keep the discussion simple, we will discuss only four
limiting cases:
\bea
(i) \ (S+P)\times (S+P)\  [(V+A)\times (V+A)] &:& 
c_1=1,\ c_2=1,\ c_3=1,\ c_4=1;
\nonumber\\
(ii)\  (S+P)\times (S-P)\  [(V+A)\times (V-A)] &:& 
c_1=1,\ c_2=1,\ c_3=1,\ c_4=-1;
\nonumber\\
(iii)\  (S-P)\times (S+P) \ [(V-A)\times (V+A)] &:& 
c_1=1,\ c_2=-1,\ c_3=1,\ c_4=1;
\nonumber\\
(iv) \ (S-P)\times (S-P) \ [(V-A)\times (V-A)] &:& 
c_1=1,\ c_2=-1,\ c_3=1,\ c_4=-1.
\eea
This choice is preferred since the $1-(+)\g_5$ projects out the weak doublet
(singlet) quark field. For the doublet fields, to maintain gauge invariance,
one must have an SU(2) partner interaction, {\em e.g.}, $\bar{s}(1-\g_5)s$
must be accompanied by $\bar{c}(1-\g_5)c$. No such argument holds for the
singlet fields.

We have chosen the interaction in a singlet-singlet form under SU(3)$_c$.
The reason is simple: one can always make a Fierz transformation to the
local operator to get the octet-octet structure. Since no spin-2 mesons
are involved, the transformations are particularly simple, as we can 
neglect the tensor terms. 

The amplitudes for the various decay processes under consideration take
the following form:

(i) $B\to\phi K$ (the contributions are same for neutral and charged channels):
\bea
i{\cal M}_s &=& ih_se^{i\phi_s} \left( -\frac{1}{2N_c}\right)\frac{1}{2}
(c_1c_3-c_2c_4) A_\phi,\nonumber\\
i{\cal M}_v &=& ih_v e^{i\phi_v}\left[ c_1c_3\left(1+\frac{1}{2N_c}
\right) + \frac{1}{2N_c} c_2c_4\right] A_\phi,
   \label{bphikampl}
\eea
where
\be
A_\phi = 2f_\phi m_\phi F_0^{B\to K}(m_\phi^2) (\epsilon_\phi.p_B).
\ee

(ii) $B\to \eta^{(')} K$ (again, since there is no $b\to s\bar{u} u$ new 
physics (NP) contribution, charged and neutral channels are equally affected):
\bea
i{\cal M}_s &=& -ih_se^{i\phi_s} \left[c_2c_3 R(P,b,s) -
\frac{1}{4N_c}\left\{(c_2c_3-c_1c_4)+(c_2c_3+c_1c_4)R(P,b,s)\right\}\right] A_P,
\nonumber\\
i{\cal M}_v &=&-ih_ve^{i\phi_v} \left[c_2c_3 +
\frac{1}{2N_c}\left(
(c_2c_3+c_1c_4) - (c_2c_3-c_1c_4) R(P,b,s) \right)\right] A_P,
\eea
where
\be
R(P,b,s)=m_P^2/2m_s(m_b-m_s)
\ee
($P=\eta,\eta'$) is the chiral enhancement factor, and
\be
A_P=if_P(m_B^2-m_K^2) F_0(q^2), \ \ q=p_B-p_K.
\ee

(iii) $B\to \eta^{(')} K^\ast$: This follows a path similar to that
discussed in (ii); however, only the axial-vector current contributes,
and the expression for $A_P$ looks slightly different:
\be
A_P=2m_{\overline{K^\ast}} f_P \left( \epsilon^\ast.p_B\right) A_0(q^2).
\ee
The chiral enhancement factor is given by
$S(P,b,s)=-m_P^2/2m_s(m_b+m_s)$.

\section{Result}

\begin{figure}[htbp]
\vspace{-10pt}
\centerline{\hspace{-3.3mm}
\rotatebox{-90}{\epsfxsize=6cm\epsfbox{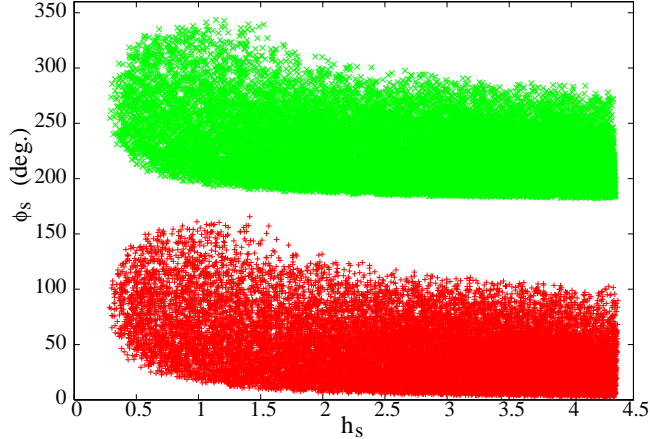}}}
\caption{Allowed parameter space for $(S+P)\times (S+P)$ (lower red) and
$(S-P)\times (S-P)$ (upper green) 
type new physics at $2\sigma$ confidence level. 
The NP amplitude is to be multiplied by $10^{-8}$.}
\end{figure}
\begin{figure}[htbp]
\vspace{-10pt}
\centerline{\hspace{-3.3mm}
\rotatebox{-90}{\epsfxsize=6cm\epsfbox{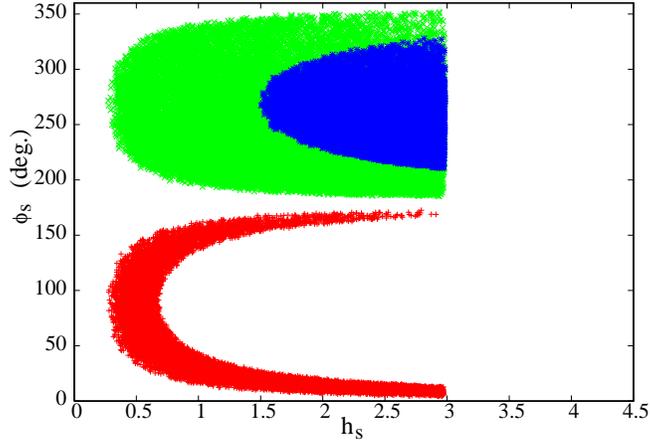}}}
\caption{Allowed parameter space for $(S+P)\times (S-P)$ (lower red) and
$(S-P)\times (S+P)$ (upper green) type new physics at 
$2\sigma$ confidence level. Also shown is the $1\sigma$ level parameter space
(upper blue) for $(S-P)\times (S+P)$ type. 
The NP amplitude is to be multiplied by $10^{-8}$.}
\end{figure}
\begin{figure}[htbp]
\vspace{-10pt}
\centerline{\hspace{-3.3mm}
\rotatebox{-90}{\epsfxsize=6cm\epsfbox{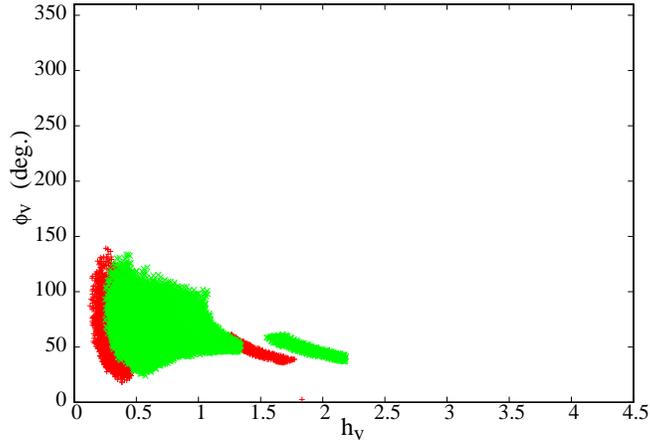}}}
\caption{Allowed parameter space for $(V+A)\times (V+A)$ (red) and $(V+A)
\times (V-A)$ (green) type new physics at $2\sigma$ confidence level. 
Note that at the bottom left corner, there is a large overlap between these
two parameter spaces.
The NP amplitude is to be multiplied by $10^{-8}$.}
\end{figure}

We are now ready to discuss our results. They are shown in figures 1-3
as the allowed parameter spaces for different Lorentz structures. The
theoretical inputs have been summarised in the previous section,
while the experimental inputs (BRs and CP asymmetries) are all listed in
table 1. Note that since we are interested only in the $\bsss$ transition,
no other decay modes (like $B\to\pi\pi$ or $B\to\pi K$)
have been taken into consideration. Although theoretically $b\to c\bar{c} s$
modes (like $B\to J/\psi K_S$) can be affected by the NP dynamics, the
justification for not including them in the analysis (which essentially
means neglecting the NP amplitude in the decay) comes {\em a posteriori}
from the allowed amplitude of NP; this is so small compared to the tree-level
Cabibbo-allowed SM amplitude that one can safely disregard them.

Apart from the BRs and CP asymmetries, we also use the following results from
\cite{hfag}:
\begin{itemize}
\item
$\sin(2\beta)$ from charmonium modes: $0.725\pm 0.037$;
\item 
$\sin(2\beta)$ from the fit to the Unitarity Triangle plus $\varepsilon_K$
\cite{utfit}: $0.729 \pm 0.042$
\item 
$\sin(2\beta)$ from $\bsss$ transitions: $0.43\pm 0.07$;
\item
$41^\circ < \gamma < 68^\circ$; dependence on $\gamma$, coming through
the SM amplitudes of $B\to\eta^{(')}$ modes, is completely negligible.
\end{itemize}
The magnitude of the CKM elements are taken from \cite{pdg04}, with only
the unitarity constraint imposed. All elements except $V_{ub}$ and $V_{td}$ 
are taken to be at their central values, while the full uncertainties for
these two elements were incorporated. Again, due to the nature of the 
transition we are considering, this variation has almost no effect.

With the expressions for BRs and CP asymmetries at hand, we perform a scan
on the new physics (NP) parameters $h_{s(v)}$ and $\phi_{s(v)}$. If we 
assume the NVQM to hold for the SM amplitude of $B\to\eta(')$, no
common region is found that satisfy all the data. This leads us to
relax the BR constraint on the $B\to\eta(')$ channels, {\em i.e.},
we assume that there is some SM dynamics beyond NVQM that generates
a significant fraction of the amplitudes. We also assume that this
dynamics does not affect the CP asymmetry predictions, {\em i.e.},
in the absence of NP, the mixing-induced CP asymmetry in $B\to
\eta' K_S$ should give $\sin(2\beta)$ and the direct CP asymmetries
in all these channels should vanish. The implementation of such a
logic is performed by demanding that the contribution of the SM amplitude
(NVQM part) plus that of the NP should not overshoot the experimental
BRs (we find that the beyond-NVQM dynamics should generate 30-40\% of
the decay amplitudes for these modes). 
 
Only the regions that satisfy all constraints are shown in figures 1-3. 
For the scalar-type new physics, it can be seen from figures 1 and 2 
that all Lorentz structures are allowed
when we take the experimental data at the $2\sigma$ confidence
level (CL), and their ranges are also in the same ballpark, but what differs is
the weak phase. This behaviour can easily be explained from the expressions
of the NP amplitudes. Since we do not consider $SU(2)$ conjugate channels,
there is no way to discriminate, as far as the magnitudes are concerned,
between $SU(2)$ singlet and $SU(2)$ doublet type interactions.
We have not taken into account any strong phase difference between the
NP amplitude and that of the SM; since the direct CP asymmetries are compatible
with zero for almost all channels (except $B\to\eta K$ by about $1\sigma$) 
this is not such
a drastic assumption. We have checked that if we switch the strong phase on,
even the $B\to\eta K$ CP asymmetry data can easily be accomodated. 
However, strong phase differences cannot alleviate the necessity of
introducing beyond-NVQM dynamics. 
If we take the experimental errors at $1\sigma$ instead of $2\sigma$,
only the $(S-P)\times (S+P)$
Lorentz structure remains allowed.

It is easy to understand why a simple NP model, without any contribution
from beyond-NVQM dynamics, cannot explain the BR anomalies
of the $\eta$ and $\eta'$ modes. If we take only a $\bsss$ structure,
the NP amplitude contributes equally to both $B\to\eta'$ and $B\to\eta$,
the only difference being the factors $f^s_\eta$ and $f^s_{\eta'}$, which
happen to be numerically close. Thus, if the NP amplitude is taken to be
so high as to jack up the $B\to\eta' K$ BR to the experimental value, it
also raises the $B\to\eta K$ BR, which, however, is expected to be much
smaller. The weak phase being fixed, we cannot even use the argument of
constructive interference with the SM in one case and destructive interference
in the other. Thus, one is forced to admit the possibility of some
SM dynamics beyond NVQM to explain these decays.

For the vector channels, only two Lorentz structures are allowed:
$(V+A)\times (V+A)$ and $(V+A)\times (V-A)$, both at $2\sigma$.
The structure $(V-A)\times (V+A)$ is a near miss; one can have a tiny
overlap if the data (particularly the CP asymmetry data) shifts a bit.
Thus, it is not entirely prudent to rule out such a structure right now.
However, if the experimental
error bars are taken at $1\sigma$, all the allowed structures get ruled
out. So, if the numbers remain as they are, and the error bars get 
smaller, there will be only one possible Lorentz structure of new physics.
However, it is too early for such a prediction.

There are many well-motivated models that can generate such Lorentz structures.
For example, minimal Supersymmetry can produce such vector-axial vector
interactions, so does the model with extra gauge bosons in the left-right
symmetric models. R-parity violating supersymmetry can produce tree-level
scalar type interactions. 

We have also checked the robustness of our results with the variation of
the regularisation scale from $\mu=m_b/2$ to $\mu=m_b$, and found it to 
be minimal. This is not unexpected since at the NLO level the scale dependence
of the Wilson coefficients becomes small. 
Finally, we have found that with the variation of $N_c$, some regions that
are disallowed for $N_c=3$ become allowed, but the result is more or less 
stable. For the scalar case, the results are absolutely $N_c$-stable (with
a slight variation of the parameter space),
when we vary $N_c$ over the range 2 to 4. All structures are allowed at
$2\sigma$ CL but only one, $(S-P)\times (S+P)$,
 at $1\sigma$ CL. For $N_c=\infty$, there
are no allowed regions for any Lorentz structure, but that is only to be
expected, since the $B\to\phi K$ NP amplitude vanishes at that limit.

For the vector case, with $N_c=2$, the $(V+A)\times (V+A)$ and
the $(V+A)\times (V-A)$ parameter spaces that we got for $N_c=3$ remain
almost
unchanged. The other two structures now get allowed over a tiny parameter 
space, namely, $h_v<0.4\times 10^{-8}$ and $\pi < \phi_v < 2\pi$. The reason
is that with the lowering of $N_c$, the $c_2$-dependent term in eq.\ 
(\ref{bphikampl}) becomes more important. These
results are true for $2\sigma$ CL. Only the first two remain at $1\sigma$ CL, 
over the range $0.3\times 10^{-8} < h_v < 1.1\times 10^{-8}$, 
$20^\circ < \phi_v < 100^\circ$ (for $(V+A)\times (V+A)$), and
$1.2\times 10^{-8} < h_v < 1.6\times 10^{-8}$, 
$28^\circ < \phi_v < 40^\circ$ (for $(V+A)\times (V-A)$). For $N_c=4$
the results are almost identical with $N_c=3$. The reason for this is
essentially the highly unstable nature of the $B\to\phi K$ BR with
the variation of $N_c$ (remember that the variation of $N_c$ is not 
the same as keeping $N_c=3$ and taking different models of mesonic
decays).  

The result is stable with the variation of the other parameters (quark masses,
form factors, CKM elements, etc.). 

We have not discussed here the polarisation data of $B\to VV$ ({\em e.g.},
$B\to\phi K^\ast$) channels. This will be treated in detail in a subsequent
publication \cite{nandi3}. 
We would just like to point out that a scalar-type new physics
may significantly reduce the longitudinally polarised fraction in the decay
$B\to\phi K^\ast$ \cite{das} (for the experimental data see \cite{hfag}). 

\section{Signatures in \boldmath$B_s$ \unboldmath System} 

We expect a large number of $B_s$ decay channels to be probed in the
hadronic machines. The first ones will include the gold-plated channel
$B_s\to J/\psi\phi$ and $B_s\to D_s^+ D_s^-$, both occurring through the
tree-level decay $b\to c\bar{c} s$. In the SM, the mixing-induced CP
asymmetry is expected to be very small. It has been pointed out that
new physics may introduce a nonzero CP asymmetry.
There are two sources for that: (i) a new amplitude in the $\bsbsbar$
mixing, with a nonzero weak phase, and (ii) a new contribution in the
$b\to c\bar{c} s$ decay. It may so happen that both of them are
present, and one needs a careful disentanglement of the two effects
\cite{gg-ak-ad}. 

It can be shown that the new physics amplitude, given by $h_{s(v)}$, is at most
10\% compared to the corresponding SM quantity for $b\to c\bar{c} s$ decay,
namely, $(G_F/\sqrt{2}) V_{cs}V_{cb}$. This, as far as the decay is concerned,
NP effects will be extremely hard, if not outright impossible, to detect.
This limits our discussion to NP effects in $\bsbsbar$ mixing only. Note that
this is not true if we consider final states like $\phi\phi$ or 
$\eta(')\eta(')$ that proceed through $\bsss$ transition. 

Unlike the decay, the calculation of the mixing amplitude depends on the
precise structure of the NP Lagrangian, and one must perform a model-dependent
study. Let us take the $1\sigma$ result, {\em i.e.}, the only possible
Lorentz structure is $(S-P)\times (S+P)$. One of the prime candidates to
generate such a structure is R-parity violating (RPV) supersymmetry, which
leads to a slepton-mediated amplitude proportional to $\lambda'_{i22}
{\lambda'_{i23}}^\ast$. It can easily be seen that
\be
h_s=-\frac{\lambda'_{i22}{\lambda'_{i23}}^\ast}{4m_{\tilde\nu_i}^2}.
\ee
Thus, to be consistent with the decay data, the magnitude of the product
of these two $\lambda'$s should be somewhere between 
$0.6\times 10^{-3}$ and $1.2\times 10^{-3}$,
 and the associated weak phase between 0 and $\pi$ (the minus sign takes care
of a phase factor of $\exp(i\pi)$) for
a sneutrino mass of 100 GeV. The necessary formalism to evaluate RPV
contributions to the neutral meson mixing amplitude has been worked out
in \cite{gg-arc,ak-jps3}. Following these and taking $\Delta M_s>14.4$ ps$^{-1}$
one can calculate the effective CP-violation in $\bsbsbar$ mixing.

The result, however, is not really encouraging. The reason is simple:
the SM amplitude is too large, enhanced from the $\bbbar$ mixing amplitude
by at least a factor of $|V_{ts}|^2/|V_{td}|^2$ times
$\xi^2$, where $\xi=f_{B_s}\sqrt{B_{B_s}}/f_{B_d}\sqrt{B_{B_d}}
=1.21\pm 0.06$ \cite{0406184}.
Furthermore, only a lower
bound on the SM amplitude exists. The NP amplitude with such a weak
coupling as obtained from the decay fit cannot compete with the SM
amplitude. We find $\sin(2\beta_s)$, the effective phase from the
$\bsbsbar$ box, to be never greater than $0.1$. Thus, it will be extremely
unlikely to have a signal of this type of NP in $B_s\to J/\psi\phi$ decay.
Note that for such a Lorentz structure, there will be a small but
nonzero NP contribution to the $b\to c\bar{c} s$ decay.

One can, of course, integrate out the left-handed squark field and get
the decay $B_s\to\ell_i^+\ell_i^-$ (this is a strongly model-dependent
statement and is true only in the context of RPV supersymmetry).  
For squarks at about 300 GeV, leptonic decay widths of $B_s$ will be 
greatly enhanced; in
particular $B_s\to\mu^+\mu^-$ will be close to the experimental limit.
(This set of couplings also enhances $B\to K^{(\ast )}\ell^+\ell^-$ 
decay widths, but the constraints are more model-dependent.)
Thus, these channels will be worth watching out. 



{\em Acknowledgements:}
A.K. thanks the Department of Science and Technology, Govt.\ of India,
for the project SR/S2/HEP-15/2003. S.N. and J.P.S. thank UGC, Govt.\ of
India, and CSIR, Govt.\ of India, respectively, for research fellowships.

\end{document}